# Pressure-Driven Fermi Surface Reconstruction of Chromium

R.L. Stillwell [1,2], D.E. Graf [2], W.A. Coniglio [2], T.P. Murphy [2], E.C. Palm [2], J.H. Park [2],

D. VanGennep [1,2], P. Schlottmann [1,2], S.W. Tozer [2]

[1] *Department of Physics, Florida State University, Tallahassee, FL, 32306, USA*

[2] *National High Magnetic Field Laboratory, Florida State University, Tallahassee, FL, 32310, USA*

We have observed a massive reconstruction of the Fermi surface of single crystal chromium as a function of high pressure and high magnetic fields caused by the spin-flip transition, with multiple new orbits appearing above 0.93 GPa. In addition, some orbits have field-induced effective masses of ~0.06-0.07 $m_e$, seen only at high magnetic fields. Based on the temperature insensitivity displayed by the oscillation amplitudes at these frequencies, we attribute the orbits to quantum interference rather than to Landau quantization.

## I. Introduction

The modern condensed matter community has been interested in chromium since Bridgman showed an anomalous resistivity versus temperature relationship in 1932 [1]. Over 30 years passed before this transition was confirmed to be an antiferromagnetic (AFM) transition at 311 K with an incommensurate spin density wave and charge density wave (SDW and CDW, respectively) structure below the transition by two separate groups [2,3]. They found that the wave vector is given by $\mathbf{Q} = 2\pi/a\,(1\pm\delta)$ directed along a principle axis of the body-centered cubic unit cell, where $\delta$ is the amount of incommensurability with the lattice. In addition, there was found to be a spin polarization flip at 123 K from perpendicular to parallel with the SDW vector $\mathbf{Q}$ [4]. More recently, chromium has sparked interest as a model system for studying more complex phenomena associated with quantum criticality[5,6]. Several groups have studied the effect of vanadium substitution, $Cr_{1-x}V_x$, on suppressing the AFM transition to study the quantum critical point.[7,8] Feng et al.[9] used X-ray diffraction to study the effects of pressure up to 7.2 GPa on the Fermi surface of chromium as it approached the quantum critical regime. Combining their pressure data with previous data for chemical substitution, they found that both parameters suppressed the AFM transition at the same rate until the doping concentration $x$ reached $x \leq 2.5\%$. From this they concluded that doping was responsible for suppressing $T_N$ due to disorder but that there was conversion between pressure and chemical doping of $dP/dx = 1.99$ GPa/% up to around $x \leq 2.5\%$ or $P=5$ GPa. Pedrazzini and Jaccard[10] were able to completely suppress the AFM state by critical pressure $P_c \sim 10$ GPa in measurements of electrical resistivity. Following this paper, Jaramillo et al.[11] showed that by measuring both resistivity and x-ray diffraction of the CDW, the $P_c$ was 9.71 GPa. They showed that chromium moves from a Bardeen-Cooper-Schrieffer-like ground state for $P < 7.0$ GPa into a state that is dominated by quantum fluctuations as the AFM state is suppressed. Up to this pressure the $Q$ vector does not change significantly with respect to the crystal lattice $a$ ($\Delta Q/Q=1\%$, $\Delta a/a=1.5\%$), and suggests that the Fermi surface is not changing to a large extent during the pressure suppression of the AFM state[12].

A more direct measurement of the Fermi surface is by the de Haas-van Alphen (dHvA) or Shubnikov-de Haas (SdH) technique, both of which are directly related to the band structure of the material[13]. It was recognized theoretically by Lomer[14] that the SDW is stabilized in the AFM state due to the almost-perfect three-dimensional nesting of the hole octahedron at $H$ with the electron octahedron at $\Gamma$ in the paramagnetic (PM) Fermi surface of chromium. Lomer proposed a model, shown in Fig. 1 (b), for the AFM state of chromium wherein the PM Fermi surface is translated by $\pm n\boldsymbol{Q_{SDW}}$, where $n$ is an integer, producing many overlapping orbits, a selection of which are illustrated in Fig. 5, and whose corresponding frequencies are listed in Table 1.

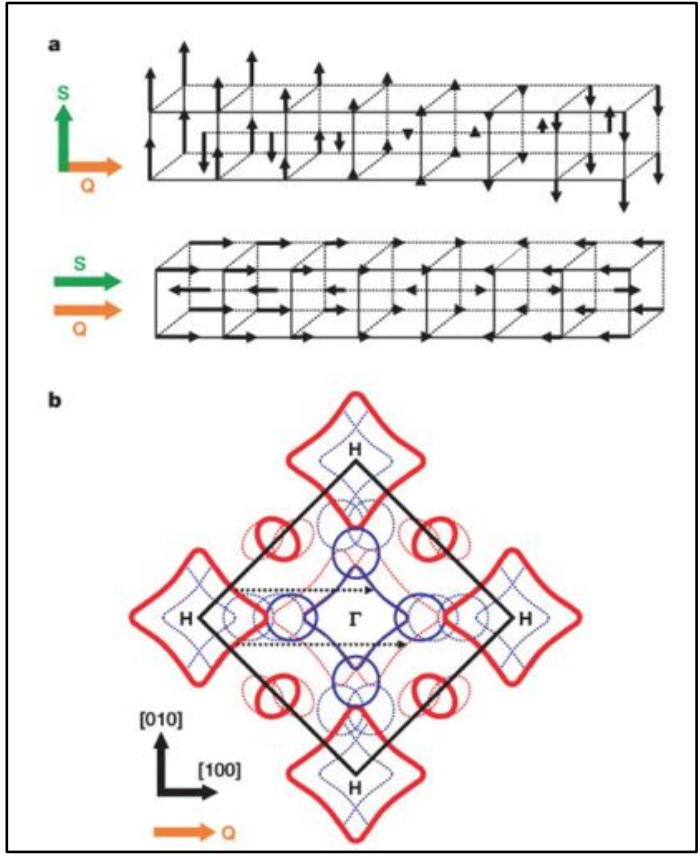

FIG. 1. **a.** Illustration of the transverse polarized state (top) above 123 K, and the longitudinal polarized state (bottom) below 123 K, with the transition temperature as stated for ambient pressure and zero magnetic field. **b.** The PM Fermi surface of chromium in the <001> plane (solid lines). The black line is the first Brillouin zone (BZ) boundary. The black dotted line shows translations of the PM Fermi surface by $\pm n\boldsymbol{Q}_{SDW}$ and illustrates the nesting of the $H$ and $\Gamma$ surfaces. The red and blue dotted lines show the translated orbitals in the BZ. All of the Fermi surfaces measured by SdH or dHvA are from the N hole ellipsoids at the edge of the BZ and orbits that can be formed by MB between overlapping, translated ellipsoids. See figure 5 for selected MB orbits allowed by reflection and transmission in the Lomer model Both figures are taken from ref. [15].

Graebner and Marcus[16] observed a rich spectrum in their measurements of the dHvA effect in chromium, see Table 1 for a comprehensive listing of reported orbits. They were able to confirm many of the orbits proposed by the Lomer model as well as confirm that, presumably as a result of the complete three dimensional nesting of the Fermi surfaces at $H$ and $\Gamma$, neither of these large octahedra was observable. Indeed, all closed orbits that have been previously reported are due to the hole ellipsoids at $N$ and the magnetic breakdown (MB) orbits created by the translations through $\pm n Q_{SDW}$ [17,18]. In contrast to the large amount of orbits observed in the dHvA paper, the oscillatory magnetoresistance (SdH) measurements by Arko et al.[19] gave a spectrum of only 7 frequencies, as opposed to 20 frequencies observed in dHvA, and only one corresponded to those found in the dHvA spectrum. Another difference between the two studies was that the SdH oscillations showed a much weaker temperature dependence of the amplitude of the oscillations. Reifenberger et al. reproduced the SdH results and concluded that the amplitude of the oscillations was "essentially independent of temperature" between 1.2 and 4 K for fields between 4 and 10.5T [20]. They were able to show that, due to MB on the $N$ ellipsoid chains, there are open orbits in the direction of $Q$ and that these orbits meet the requirements to produce QI oscillations[20].

In this article we present measurements of quantum oscillations in the surface conductivity of chromium at higher pressures and higher magnetic fields than previously measured that show a massive Fermi surface reconstruction for pressures above 0.93 GPa. This effect is reproducible with pressure cycling and has been seen in multiple samples. Quantum oscillations observed following this Fermi surface reconstruction show orbits characterized by Landau quantization as well as orbits that are more characteristic of (QI). If treated with the traditional Lifshitz-Kosevich (LK), damping parameter, these orbits have effective masses that are in the range of 0.06-0.07 $m_e$, a factor of two to five times lighter than any effective masses previously measured in single crystal chromium. These light effective masses allowed quantum oscillations to be observed up to 70 K in magnetic fields up to 35 T. At these high temperatures, the rich fast Fourier Transform (FFT) spectrum at high pressures shows multiple orbits that are not well described by LK formalism.

## II. Experimental Details

Measurements were performed on a single crystal of pure chromium[21], cut by spark erosion[22] into a cylinder with a diameter of 0.35 mm and a length of 0.45 mm, and followed by a chemical etch [23] to remove any damage caused by the electrical discharge. The sample was cut with one of the <100> axes parallel with the cylinder and oriented in the pressure cell so that **H**//<100>. All of the data presented are from one sample that was pressurized to 1.41 GPa, returned to 0.17 GPa, and then repressurized to 0.92 and 1.47 GPa. The sample was cooled at zero magnetic field through the spin polarization flip at 123 K ($T_{SF}$). Therefore, the initial state of all measurements made was with the system having an equal distribution of **Q** and spin (**S**) domains at zero pressure, see Fig. 1(a) for illustration of transverse spin polarization (top) and longitudinal spin polarization (bottom).

Pressure was applied in a piston cylinder cell (PCC) designed and manufactured at the National High Magnetic Field Laboratory (NHMFL) in Tallahassee, Florida. The cell was 45 mm long and therefore was not able to rotate in the bore of the magnet. Daphne 7474 oil was used as a pressure medium[24, 25] in a Teflon cup with an inside diameter of 3.175 mm and length of 8.76 mm. All of the measurements for the temperature dependence data were taken in a top-loading $^3$He system positioned in a 35T resistive magnet at the NHMFL. All temperatures were measured using a calibrated Cernox thermometer located ~15 mm from the sample, and both the thermometer and the sample were immersed in the $^3$He liquid/vapor. Figures 2 and 3 include data from both a Quantum Design 16 T Physical Property Measurement System (PPMS) and the 35 T resistive magnet system. All of the data presented were acquired via the tunnel diode oscillator (TDO) technique [26] with the diode at low temperatures. The TDO is an inductor-capacitor (LC) resonant tank circuit technique in which the sample under study is placed in the inductor of the circuit, and any changes in the surface conductivity of the sample are measured as a change in the resonant frequency of the circuit. To use the TDO in measurements under pressure, we placed the inductor coil inside the sample space of the PCC and pressurized the entire coil with the sample in it. The inductor coil is ~20 turns, 10 turns per layer, with an inner diameter of ~0.4 mm and height of ~0.5 mm. This results in a resonant frequency of the circuit of ~60 MHz. By incorporating the metal cell into the ground plane of the circuit, we were able to improve the signal to noise ratio by a factor of 500 to ~10 ppm. Pressure was calibrated at room temperature and then again at low temperatures (T ≈ 4 K), using the shift of the ruby R1 fluorescence peak[27]. Pressure in the cell was measured for one cooling and warming cycle, and it was verified that the pressure was stable within 1% for all temperatures measured. Hydrostaticity of the applied pressure is measured primarily by the signal to noise ratio of the quantum oscillations of the sample since crystal quality directly affects damping of the amplitude of quantum oscillations. An additional method of quantifying hydrostaticity is by the full width at half maximum (FWHM) of the R1 peak in the ruby fluorescence spectrum. The R1 peak will broaden under nonhydrostatic conditions and can therefore be monitored at high and low temperatures to see whether the sample is experiencing nonhydrostatic conditions[28]. For our measurements at P= 1.41 GPa and T= 4 K, $\Delta$ FWHM(P) =FWHM(1.41GPa)- FWHM(0)= 0.082-0.078 nm=0.004 nm=5% broadening of the R1 peak, which indicates a quasihydrostatic measurement.

## III. Results

Quantum oscillations in chromium were observed at ambient pressure up to 1.47 GPa starting at average fields near H=4T at T=2 K. Figure 2 shows background subtracted field sweeps offset for clarity. There is a distinct change in the traces above 0.92 GPa that is evident in the background subtracted traces and is also shown in the FFT spectra in Fig. 3. Nearly all of the peaks in regions 2 and 3 of Figs. 4 and 6 are seen only in the FFT spectrum at 1.47 GPa. In order to verify that these peaks were not artifacts of the data analysis procedure, several FFT window sizes were used. Within the 14-34T range used for all of the data analysis of Figs. 4 and 6, none of the peak positions changed as a function of window size. Due to the large number of orbits allowed by MB at high magnetic fields, it is difficult to identify specific orbits that are harmonics of lower-frequency orbits. Though not exact integer multiples, it is possible that the orbit identified as $\pi$ could be equal to $2\theta'$, the low frequency orbit $\varepsilon$ could either be $2\omega$ or $F_o$ and $2\zeta_2$ would be under $\theta'$ and therefore difficult to identify, though it may contribute to the amplitude of the $\theta'$ orbit. The peaks at 2500 T could be harmonics of $2\lambda$, and the largest frequency in the 1.47 GPa spectrum at 1750 T could have its harmonic in the large doublet near 3400 T. Few peaks are exactly equal to twice a lower frequency and have a proportional damping due to being a second- or third-order term in the LK formula, which leads us to believe that the nature of the complex spectrum above 0.92 GPa is not simply due to an increase in the harmonic content of the FFT. The most likely phenomenon that is causing such a significant change in the spectrum is the suppression of the longitudinal phase of the spin polarization and return to the transverse spin polarized state. The low-pressure spectra are similar to the zero-field-cooled, pure chromium samples in Ref. [7]. Differences in the spectra may also be due to slight variations in the angle of the sample with respect to the magnetic field direction.

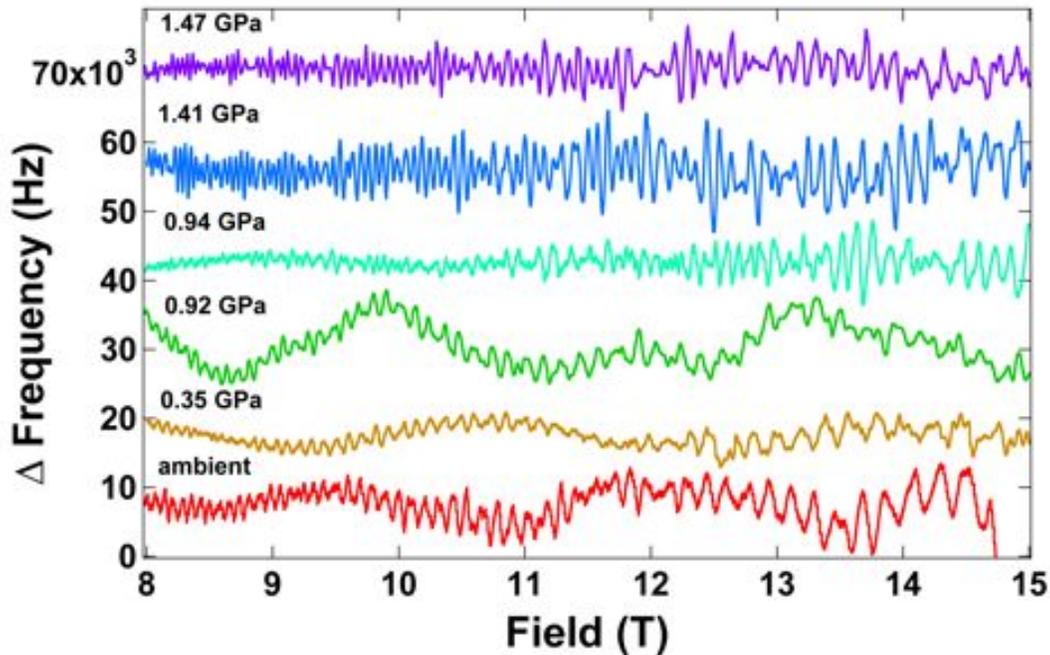

FIG. 2. (Color online) Background subtracted magnetic field sweeps with **H**//<100> at T=2 K, offset for clarity. The Fermi surface reconstruction at 0.93 GPa can be seen in the background

subtracted traces as the pressure is increased from 0.92 to 0.94 GPa. The increased number of frequencies evident in these traces is clearly seen in the FFT of this data shown in Fig. 3. All traces are the result of a fifth order polynomial subtracted from the original data. The 0.35 GPa trace was taken after releasing the pressure from 1.41 GPa to confirm reproducibility with pressure cycling. The P= 0.92 and 1.47 GPa data are from the 35 T resistive magnet; all other pressures are from the PPMS.

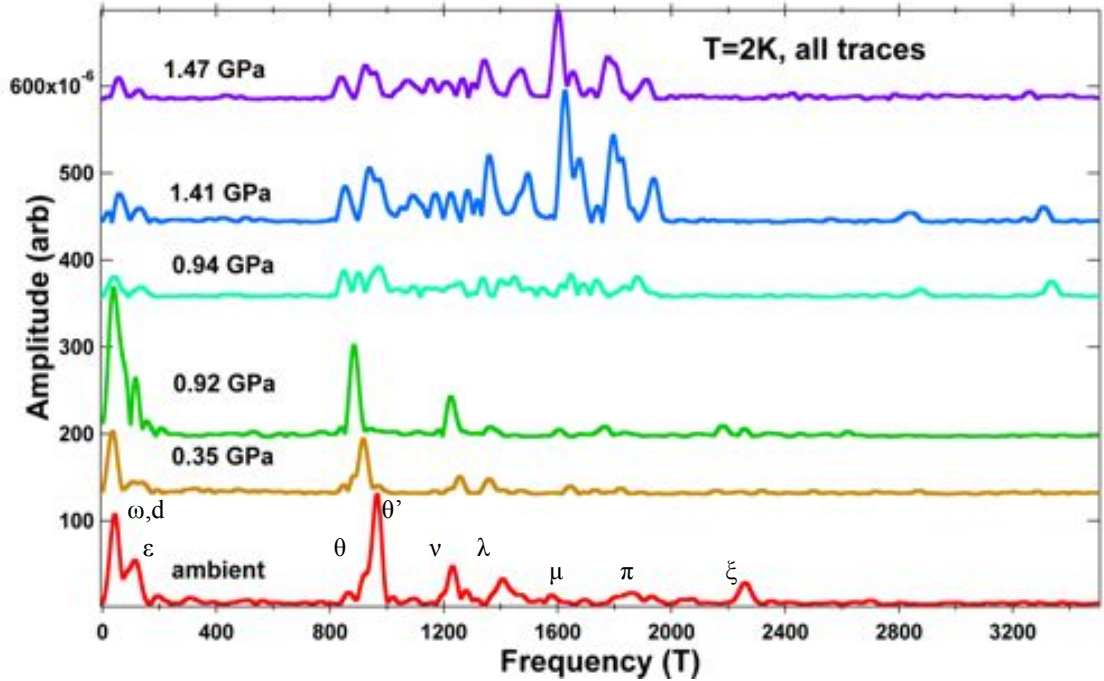

FIG. 3. (Color online) FFTs of the background subtracted data of TDO frequency vs magnetic field offset for clarity. Data show a decrease in frequency of the large peak at 910T with increasing pressure. Increasing the pressure from 0.92 to 0.94 GPa results in a Fermi surface reconstruction with the FFT spectrum showing many more orbits that are created with pressure, including higher frequency orbits at 2800 and 3300T. All FFTs were taken over a field range of 8-15T. Orbits in the ambient pressure spectrum are labeled using the conventions in Refs. [7], [17], and [29]. The 0.35 GPa trace was taken after releasing the pressure from 1.41 GPa to confirm reproducibility with pressure cycling.

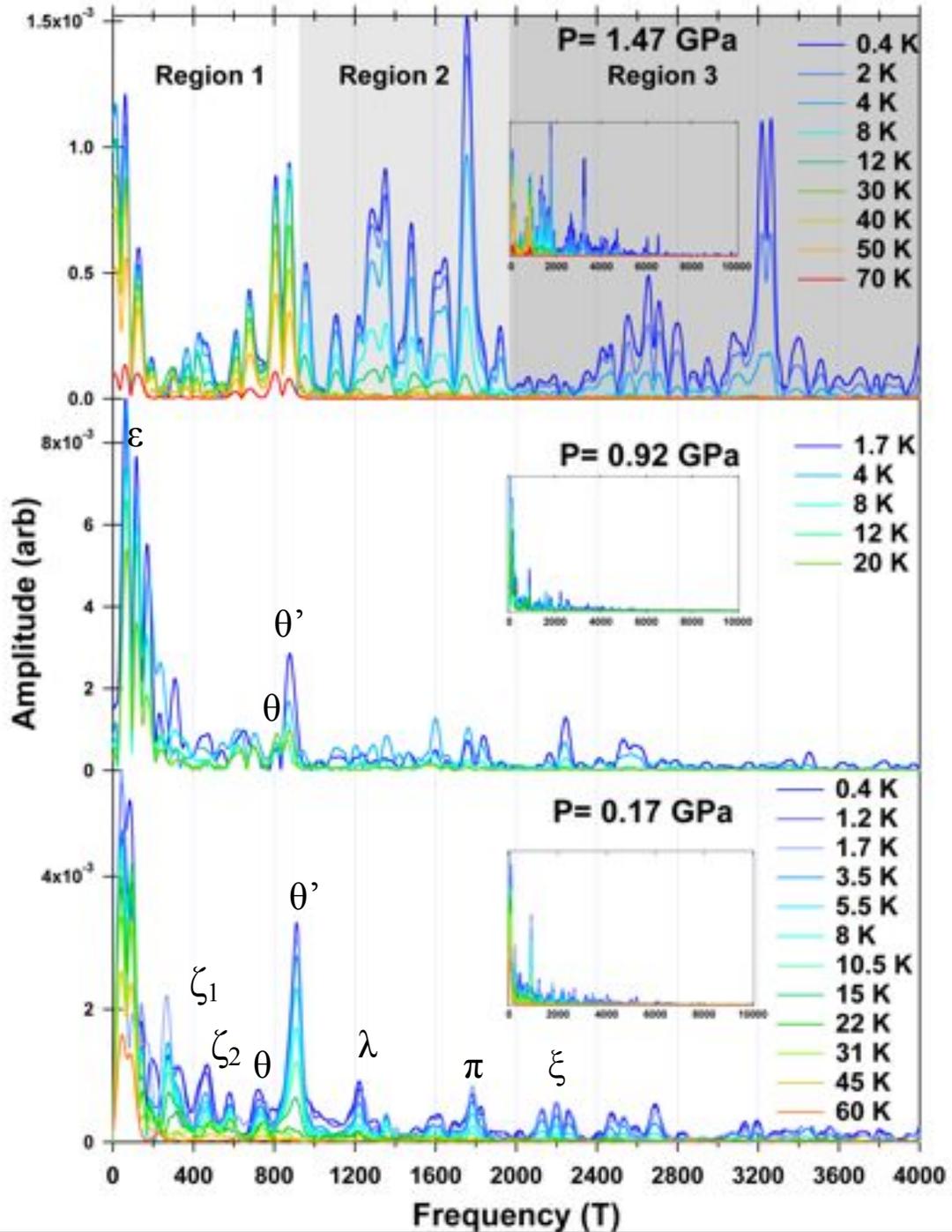

FIG. 4. (Color online) FFT spectrum development as pressure is increased from 0.17 to 1.47 GPa. All FFTs were taken in the range of 14-34T. The main panels show the largest peaks up to 4 kT, with the inset graphs showing the range out to 10 kT (vertical scale is the same for the main and inset graphs). The lowest pressure spectra show two orbits that remain to high temperatures, which we attribute to QI. At P=0.92 GPa, there are three low-frequency orbits we attribute to QI, but many of the other orbits are less distinct than they were at 0.17 GPa. After going through the Fermi surface reconstruction at 0.93 GPa, there are now three distinct regions of frequencies at P=1.47 GPa: Region 1 contains orbits that all have thermal effective masses of ~0.06 $m_e$, region

2 contains orbits in the range of ~0.4 $m_e$, similar to the orbits seen at low pressures; and region 3 contains high-frequency orbits with effective masses ~1.2 $m_e$.

| Orbit | Field Direction | Frequency (T) | | | | | | | | | | |
|---|---|---|---|---|---|---|---|---|---|---|---|---|
| | | This Work | | | Arko | | | Venema | Graebner Marcus | Vinokurova | | |
| | | 0.17 GPa | 0.92 GPa | 1.47 GPa | Field orbit direction | ambient | ambient | ambient | ambient | ambient | 0.17 GPa | 0.92 GPa |
| α | [100] | | | | | | | | | 10 | | |
| β | [100] | | | | | | | | | 30 | | |
| γ | [100] | 42 | | | e | [100] | 40 | | | 40 | | |
| | | 92 | 68 | 65 | d | [001] | 100 | | | | | |
| ε | [100] | 140 | 121 | 125 | a | [100] | 175 | | | 160 | | |
| | | | 170 | 190 | | | | | | | | |
| $\zeta_1$ | [100] | 460 | 450 | 370 | f | [100] | 250 | 424 | | 420 | 425 | 412 |
| $\zeta_2$ | [100] | 580 | 540 | 420 | | | | 564 | | 560 | | |
| | | | 625 | 610 | | | | | | | | |
| η | [100] | 725 | 700 | 675 | b | [001] | 800 | | | 845 | | |
| θ | [100] | 830 | 815 | 805 | | | | 861 | | 870 | | |
| θ' | [100] | 910 | 875 | 875 | | | | 897 | | 910 | | |
| ν | [100] | 1222 | | 955 | | | | 1259 | [001], 2680 | | 1257 | 1248 1224 |
| $\lambda_1$, N ellipsoid | [100] | | | 1105 | | | | 1331 | | 1340 | | |
| $\lambda_2$ | [100] | 1357 | | 1285 | | | | 1368 | | 1380 | | |
| μ | [001] | 1600 | 1600 | 1350 | | | | 1600 | | 1560 | | |
| n | [100] | 1700 | | 1480 | | | | 1753 | | 1740 | | |
| n' | [100] | 1780 | 1760 | 1605 | | | | 1792 | | 1804 | | |
| n'' | [100] | 1824 | 1835 | 1645 | | | | 1825 | | | | |
| ξ | [100] | 2126 | | 1755 | | | | 2181 | | | | |
| ξ' | [100] | 2200 | 2160 | 1920 | | | | 2218 | | | | |
| ξ'' | [100] | 2260 | 2245 | 2465 | | | | 2251 | | | | |
| n''' | [100] | | | 2550 | | | | 2360 | | 2400 | | |

Table 1. (Color online) Observed orbits from this paper and previous literature. Greek letters are used for all orbits except for those of Arko *et al.* (Ref. 19) who use Roman letters. This paper and Arko *et al.* are SdH; Venema *et al.* (Ref. 17), Graebner and Marcus (Ref. 16), and Vinokurova *et al.* (Ref. 30), are dHvA. More orbits are shown in Fig. 4 but are not identifiable with known orbits. The N ellipsoid is labeled $\lambda_1$; all other orbits are a result of MB between overlapping ellipsoids due to translations by $\pm n\mathbf{Q}_{SDW}$.

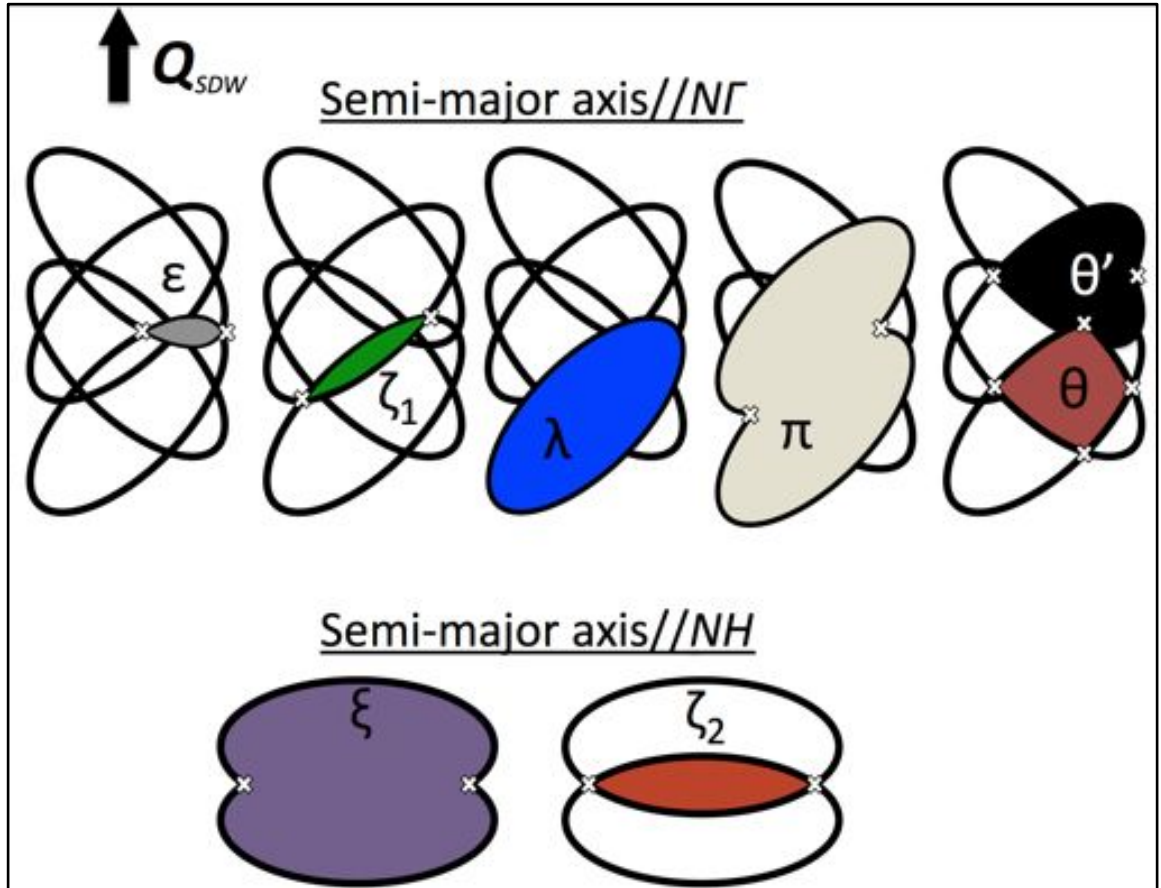

FIG. 5. (Color online) Schematic drawings of all labeled orbits in Fig. 4. The orientation of $Q_{SDW}$ for all of the orbits is shown in the figure. The magnetic field is oriented out of the page. The full ellipse at $N$ is labeled $\lambda$ (blue shading). White "x" show where transmission or reflection occurs to allow for MB orbits.

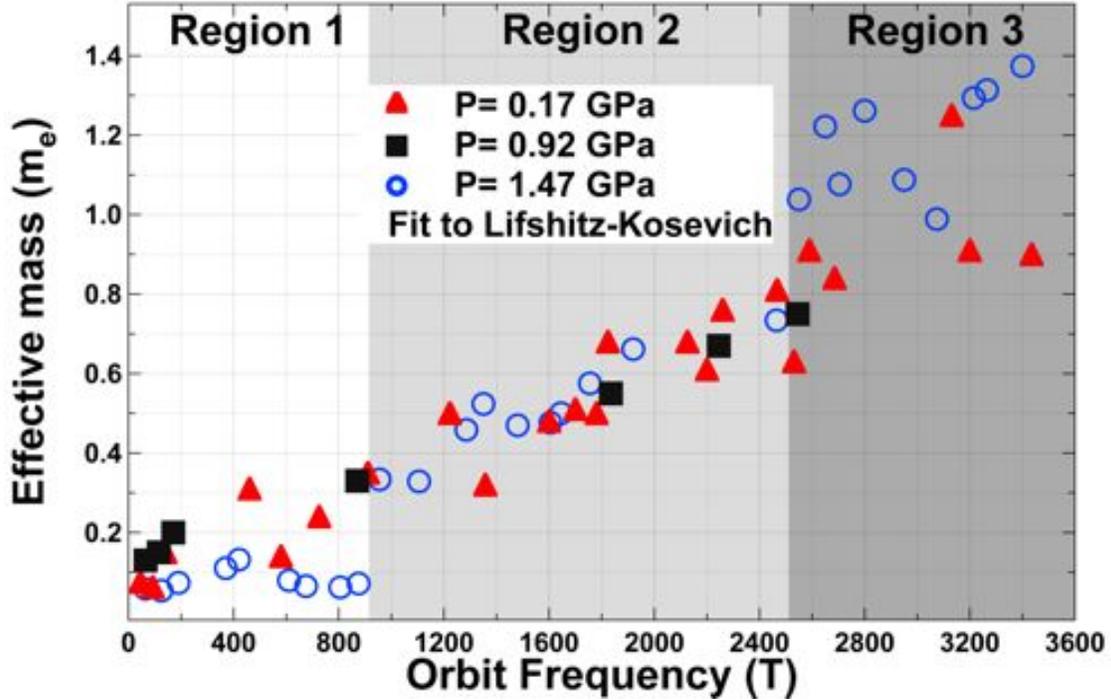

FIG. 6. (Color online) Effective masses as a function of orbit frequency. The lowest frequencies for all three pressures have effective masses that are a factor of 2-5 times lighter than any previously reported. Effective masses for the two lowest pressures follow a nearly linear dependence on frequency, whereas the P=1.47 GPa masses (circles) do not increase until ~950 T, at which point there is a discontinuous increase in mass and a change to linear dependence on frequency.

The effective masses reported for orbits in chromium at ambient pressure are between 0.18 and 0.5 $m_e$ [16]. The two previous dHvA studies under pressure did not report values for the effective masses as a function of pressure, [17, 30] and our paper examines SdH under pressure. Our SdH of pressures up to 1.47 GPa in high magnetic fields to 35 T show pressure-tuned effective masses that decrease by a factor of 2-5. The plots in Fig. 4 show temperature dependence of the FFT spectrum as pressure is increased from the lower panel to the upper panel. The lower panel at 0.17 GPa features a large peak at 910T, which we identify as the $\theta'$ orbit, with an effective mass of 0.34 $m_e$. That peak shifts to lower frequency at 0.92 GPa with an effective mass of 0.3±0.1 $m_e$. The error in the measurement of the effective mass comes from the temperature evolution of the spectrum. As the temperature is increased from 1.7 to 8K, the amplitude of the peak diminishes as would be expected from the LK formulation for quantum oscillations [13], and the effective mass is close to the low pressure value. For temperatures above 8 K the rate at which the peak amplitude is decreasing slows down until the peak at 20 K is actually larger than the peak at 12 K. This can be seen in Fig. 7 for the LK fit to the $\theta'$ orbit at 0.92 GPa, where the 20 K point is higher in amplitude and far from the LK fit line. We attribute this behavior to the Fermi surface undergoing reconstruction in this pressure range. At 1.47 GPa, the $\theta'$ peak is still one of the largest; however, the $\theta$ peak has nearly the same spectral weight as the $\theta'$ peak. Comparing the middle and upper panels of Fig. 4, one can see that as the pressure is increased, the $\theta$, $\theta'$ doublet is becoming better resolved. At higher pressure, this doublet is one of the dominant features of the FFT spectrum. It can also be seen that the amplitude of these peaks at higher pressures is roughly a third of the height of the peak at 0.17 GPa. As shown in Fig. 3, when the pressure was released from 1.41 to 0.35 GPa, the amplitude of the peaks is recovered when the pressure is released below 0.93 GPa. This verifies that the decreased FFT amplitudes at 1.47 GPa is due to the Fermi surface reconstruction and not to nonhydrostatic conditions at high pressure. LK fitting to the temperature dependence at 1.47 GPa gives effective masses of ~0.06 $m_e$ for frequencies in region 1, ~0.3-0.8 $m_e$ in region 2, and ~1.2 $m_e$ in region 3 (Figs. 4 and 6).

## IV. Discussion

The results of this experiment present two phenomena that are unexpected based on previous literature. The most striking is the temperature dependence of the amplitudes of the orbits between 200 and 950T at 1.47 GPa. The effective masses for some of the orbits found by LK fitting are 2-5 times lighter than the ambient pressure values. The other surprising finding is the Fermi surface reconstruction at 0.93 GPa. The decrease in the effective mass by a factor of 2-5 implies that either the curvature of the Fermi surface has increased by the same amount, since the effective mass is proportional to the derivative of the area with respect to energy, or that the interaction environment of the electron has changed in a significant way. The relatively small change in frequency of the $\theta'$ orbit as the pressure is increased from 0.92 to 1.47 GPa constrains the $k$-space area to a proportionally small change. This in turn requires the eccentricity of the orbit to change by a factor of 5 while maintaining the same area. Though this is possible, Feng et al.[9] measured $Q(P)$ and showed that it flattens off with pressure, which indicates that the underlying PM Fermi surface changes little as a function of pressure.

      A possible explanation for this behavior is an increasingly dominant contribution to the signal from QI oscillations. The magnetoresistance measurements of Arko et al.[31] proved the existence of open orbits along $Q$, which is required for QI orbits to be possible. Reifenberger et al.[20] showed that open orbits on the overlapping $N$ hole

ellipsoids present MB junctions enabling QI oscillations.

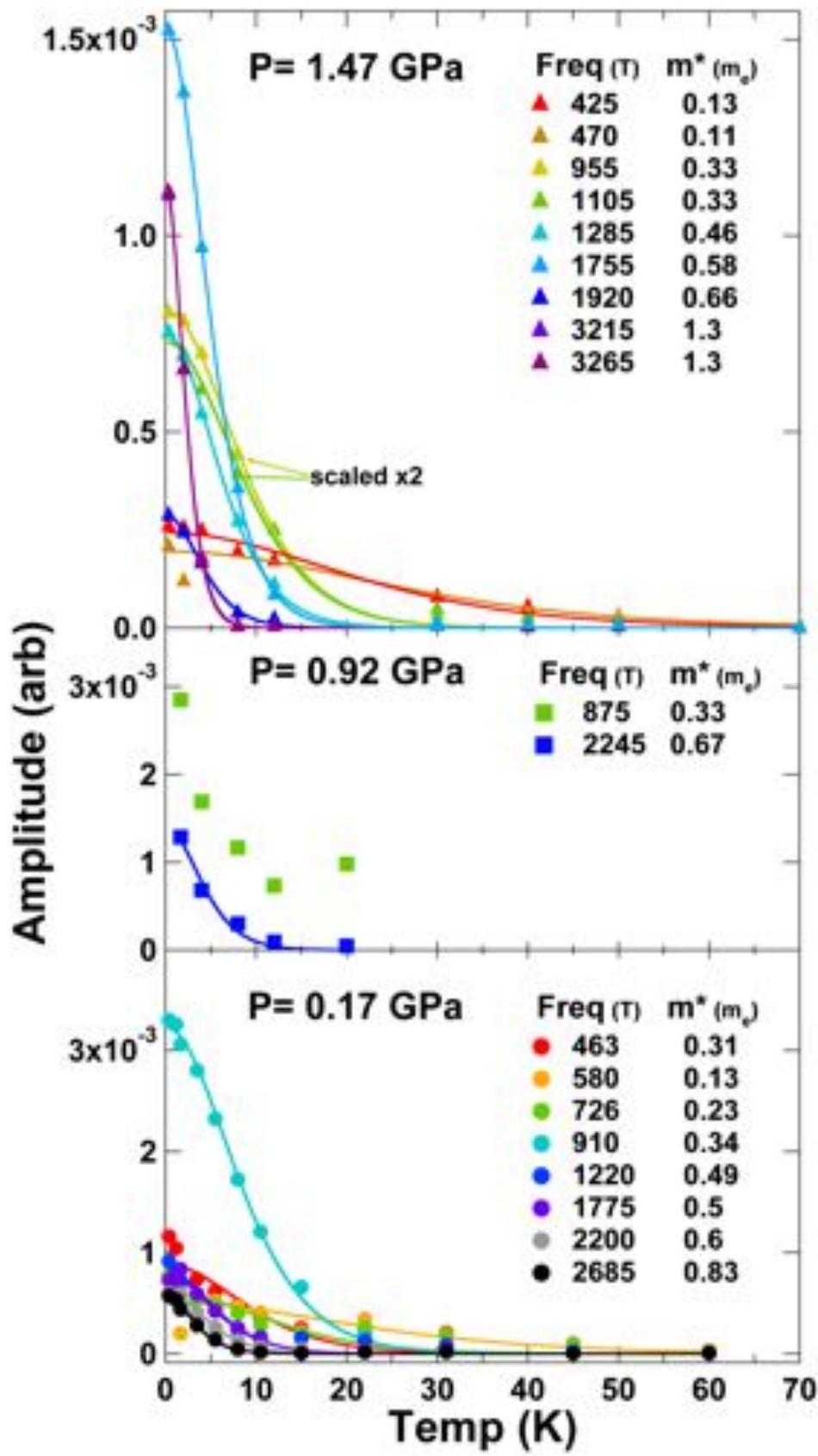

FIG. 7. (Color online) LK fits to the temperature dependence of the amplitudes of the FFT peaks for all three pressures. The traces for 955 and 1105 T at P=1.47 GPa were multiplied by 2 for clarity.

Reifenberger et al. gave a detailed treatment for two of the possible QI trajectories, only one of which, $F_o$, he was able to observe experimentally in magnetic fields up to 10.8 T at liquid helium temperatures. The Fermi surface topology of chromium allows many electron trajectories with which to create QI orbits. One of the possible QI orbits with a *k*-space area similar to $\theta'$ is illustrated in Fig. 8.

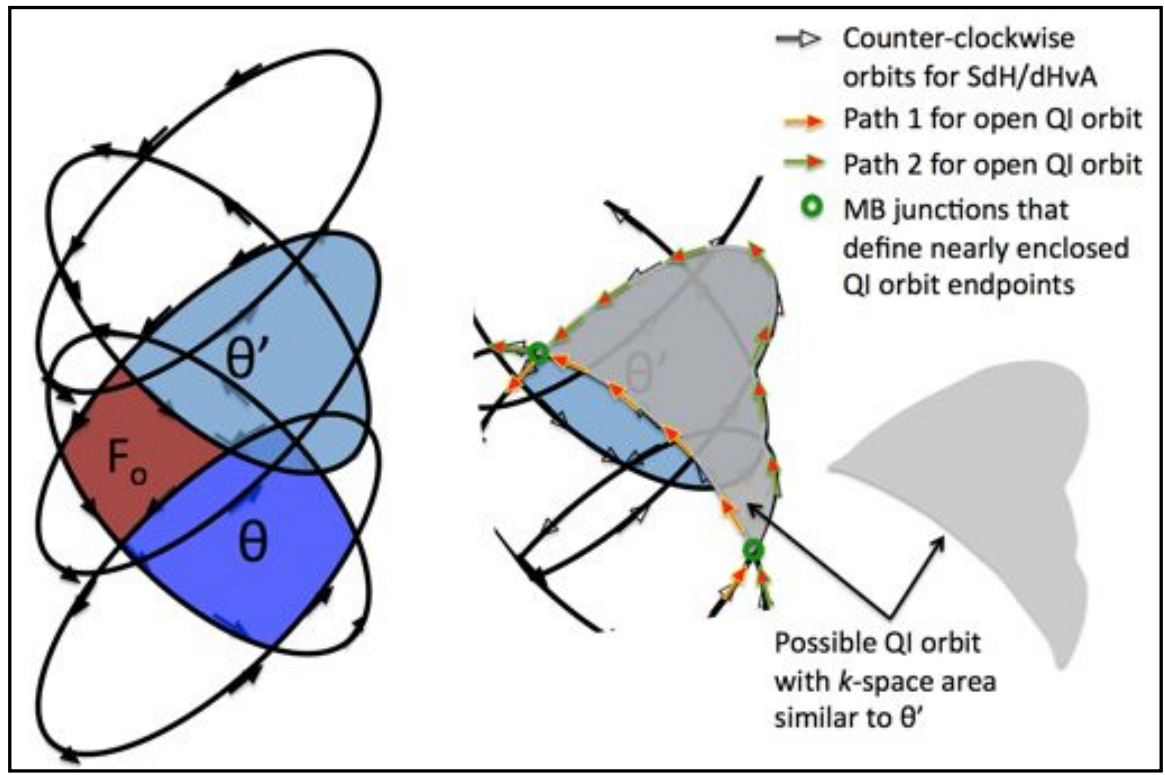

FIG. 8. (color online) (left) Illustration of the translation by $\pm n\boldsymbol{Q}$ of the hole ellipse at *N* and two of the resulting orbits, $\theta$ and $\theta'$, made possible by MB. $F_o$ is the QI orbit described in Ref. 20. The arrows show counterclockwise (CCW) rotation orbits that would allow SdH and dHvA oscillations. (middle) A portion of the same illustration shown on left but focusing on the $\theta'$ orbit with a possible QI orbit overlaid (shaded in gray). The red arrows show the nearly enclosed *k*-space area. The yellow and green outlines on the red arrows distinguish two separate, open orbits that nearly intersect at the MB junctions shown by the green circles. As shown by the black outlined arrows, both of the QI orbits follow the CCW rotation shown in the illustration at left. (right) Nearly enclosed *k*-space area of the possible QI orbit with similar area, and therefore similar oscillation frequency, as the $\theta'$ orbit.

In order to produce oscillations from QI that are proportional to 1/H, these open orbits must form nearly enclosed *k*-space areas with nodes at MB junctions. The probability *p* of transmission at these junctions is given by $p = \exp(-\Delta/H)$, $\Delta \approx (m^* E_g/E_F)$, where $E_g$ and $E_F$ are the energy gap between the two energy bands and the Fermi energy, respectively. In our experiment we increased the probability of transmission both by applying pressure, which decreases the size of the energy gap $E_g$, and by subjecting it to high magnetic fields. This leads to the state seen at the lowest pressure in Fig. 4, where only the lowest frequency orbits of ~50-150 T, labeled *ε* and *ω* in this paper and $F_o$ in Ref. 20, show the insensitivity to temperature characteristic of QI oscillations. As pressure is increased to 0.92 GPa, the system is at the threshold of the Fermi surface reconstruction. The lowest frequencies are still the only orbits that demonstrate temperature insensitivity, but many of the orbits seen at low pressure are no longer distinct. Then at 1.47 GPa, the Fermi surface has gone through reconstruction. The reconstructed Fermi surface contains many new orbits at higher frequencies, labeled as regions 2 and 3 in Figs. 4 and 6, as well as many new temperature insensitive orbits below 950T, labeled region 1 in Figs. 4 and 6. Region 1 in Fig. 6 highlights that at 1.47 GPa, the effective masses stay roughly the same up to 950 T, at which point there is a discontinuous increase in mass. Within region 2 the effective masses for all three pressures are in the same range, and all increase linearly as a function of increasing orbit frequency. Then at 2550 T there is another discontinuous increase in the effective masses for the 1.47 GPa data. This suggests that in the reconstructed Fermi surface there are sets of orbits over regions 2 and 3 that have areas perpendicular to the magnetic field that are of similar eccentricity but are only available at high pressures. These orbits follow LK temperature damping, and we attribute them to SdH oscillations. In addition, due to an increased probability of MB at high pressure and high magnetic field, we are able to detect temperature insensitive oscillations from larger *k*-space areas than previously reported that we attribute to QI. The temperature dependence of the FFT amplitude for several of these orbits is shown in Fig. 9. The solid line is a fit line to the 875T frequency using Amp~exp(-$T^3$), which would be expected for a QI orbit. The dashed line is a fit to the same data using the traditional LK damping parameter, Amp~T/sinh (T). As can be seen from Fig. 9 the fits begin to diverge above 50 K with the QI fit performing better above this temperature.

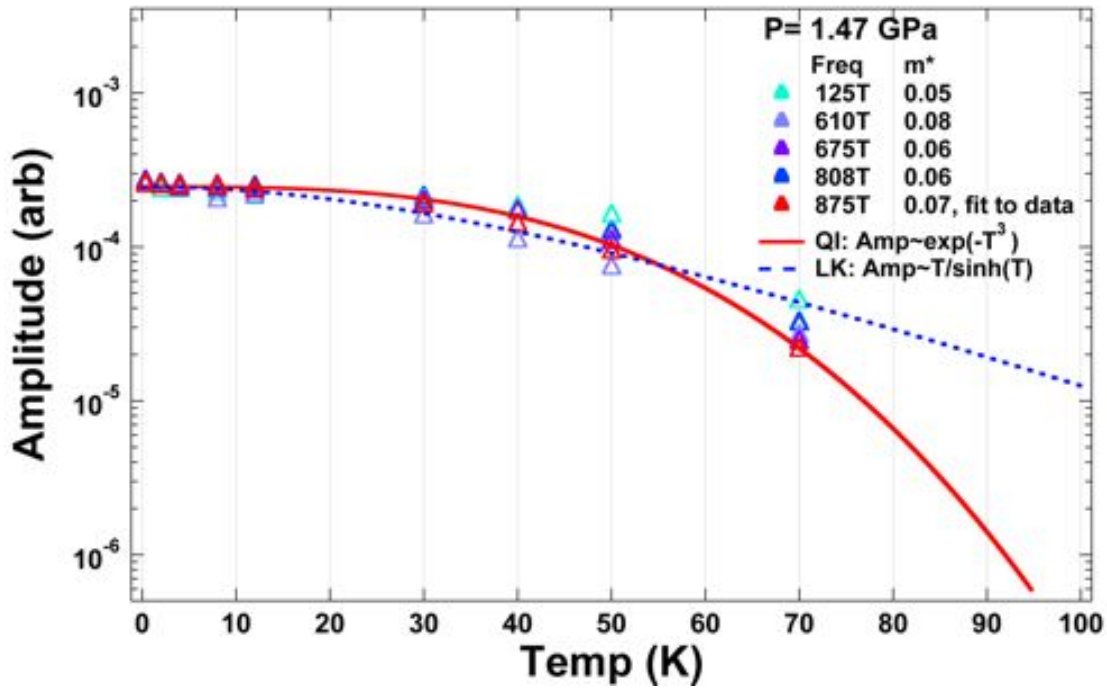

FIG. 9. (Color online) The temperature dependence of the amplitudes of the QI orbits for chromium at P= 1.47 GPa. The dashed line is a fit to the 875 T frequency data using the standard LK damping parameter, and the solid line is a fit to the same data using the semiclassical QI term. Oscillation amplitudes for orbits are normalized to the same value for the lowest temperature point. The QI fit follows the trend of all the orbits and shows the functional difference at the highest temperatures.

## V. Conclusion

We have observed quantum oscillations in chromium using quasihydrostatic pressures up to 1.47 GPa in magnetic fields to 35 T at temperatures up to 70 K. Our results at low pressure are consistent with previous studies of the Fermi surface but show a reconstruction above 0.93 GPa. We observe oscillations periodic in inverse magnetic field that give increasingly light thermal effective masses as a function of pressure and magnetic field. Based on our results, as well as previous SdH results for chromium, we attribute these light mass orbits to QI oscillations. Since our measurements are made on non-field-cooled samples measuring surface conductivity, the directional dependence required for a full QI treatment of the data is not possible. Future experiments to clearly identify whether these orbits are due to QI must be done by resistivity measurements from ambient to high pressure in high magnetic fields.

## Acknowledgements

This paper was funded by the Department of Energy (DOE) National Nuclear Security Administration under Grant No. DE-FG52-10NA29659 and No. DE-NA0001979 and performed at the NHMFL in Tallahassee, Florida, which is supported by National Science Foundation Cooperative Agreement No. DMR-1157490 and by the State of Florida. P.S. is supported by DOE Grant No. DE-FG02-98ER45707. We thank Robert Schwartz, Vaughan Williams, and Danny McIntosh for technical assistance and Y.J. Feng for useful discussions.

1. P. W. Bridgman, Proc. Am. Acad. Arts Sci. **68**, 27 (1933).
2. L. M. Corliss, J. M. Hastings, and R. J. Weiss, Phys. Rev. Lett. **3**, 211 (1959).
3. V. N. Bykov, V. S. Golovkin, N. V. Ageev, V. A. Levdik, and S. I. Vinogradov, Dokl. Akad. Nauk SSSR **128**, 1153 (1959).
4. H. Umebayashi, G. Shirane, B. C. Frazer, and W. B. Daniels, J. Phys. Soc. Jpn. **24**, 368 (1968).
5. A. Yeh, Y. A. Soh, J. Brooke, G. Aeppli, T. F. Rosenbaum, and S. M. Hayden, Nature **419**, 459 (2002).
6. M. Lee, A. Husmann, T. F. Rosenbaum, and G. Aeppli, Phys. Rev. Lett. **92**, 187201 (2004).
7. J. F. DiTusa, R. G. Goodrich, N. Harrison, and E. S. Choi, Phys. Rev. B **82**, 075114 (2010).
8. E. Fawcett, H. L. Alberts, V. Y. Galkin, D. R. Noakes, and J. V. Yakhmi, Rev. Mod. Phys. **66**, 25 (1994).
9. Y. Feng, et al., Phys. Rev. Lett. **99**, 137201 (2007).
10. P. Pedrazzini and D. Jaccard, Physica B **403**, 1222 (2008).
11. R. Jaramillo, Y. Feng, J. Wang, and T. F. Rosenbaum, Proc. Nat. Acad. Sci. USA **107**, 13631 (2010).
12. R. Jaramillo, Y. J. Feng, J. C. Lang, Z. Islam, G. Srajer, P. B. Littlewood, D. B. McWhan, and T. F. Rosenbaum, Nature **459**, 405 (2009).
13. D. Shoenberg, *Magnetic oscillations in metals* (Cambridge University Press, Cambridge, 1984).
14. W. M. Lomer, Proc. Phys. Soc. London **80**, 489 (1962).
15. R. K. Kummamuru and Y. A. Soh, Nature **452**, 859 (2008).
16. J. E. Graebner and J. A. Marcus, Physical Review **175**, 659 (1968).
17. W. J. Venema, R. Griessen, and W. Ruesink, J. Phys. F **10**, 2841 (1980).
18. E. Fawcett, Rev. Mod. Phys. **60**, 209 (1988).
19. A. J. Arko, J. A. Marcus, and W. A. Reed, Physical Review **185**, 901 (1969).
20. R. Reifenberger, F. W. Holroyd, and E. Fawcett, J. Low Temp. Phys. **38**, 421 (1980).
21. Single crystals were cut from a larger single crystal wafer bought from Alfa Aesar with a stated purity of 99.996%.
22. Micro-EDM (Electrical Discharge Machining) by Hylozoic products. Settings for the EDM were 0.05mJ for the discharge energy and 0.2mW for the peak power.
23. Chromium Etchant, standard (#651826) from Sigma Aldrich, is a ceric ammonium nitrate-based etchant.
24. K. Murata, et al., Rev. Sci. Instrum. **79**, 085101 (2008).
25. I. R. Walker, Rev. Sci. Instrum. **70**, 3402 (1999).
26. C. T. Vandegrift, Rev. Sci. Instrum. **46**, 599 (1975).
27. G. J. Piermarini, S. Block, J. D. Barnett, and R. A. Forman, J. Appl. Phys. **46**, 2774 (1975).
28. N. Tateiwa and Y. Haga, Rev. Sci. Instrum. **80**, 123901 (2009).
29. D. W. Ruesink and I. M. Templeton, J. Phys. F **14**, 2395 (1984).
30. L. I. Vinokurova, A. G. Gapotchenko, E. S. Itskevich, E. T. Kulatov, and N. I. Kulikov, JETP Lett. **30**, 246 (1979).
31. A. J. Arko, J. A. Marcus, and W. A. Reed, Physical Review **176**, 671 (1968).